\documentclass{article}
\usepackage{amsfonts}

\def\genus{\mathrm{genus}}
\def\Prym{\mathrm{Prym}}
\def\ind{\mathrm{ind}}
\def\Ker{\mathrm{Ker}}
\def\Coker{\mathrm{Coker}}

\newtheorem{proposition}{Proposition}
\newtheorem{theorem}{Theorem}

\begin{document}

\author{Iskander A. TAIMANOV
\thanks{Institute of Mathematics, 630090 Novosibirsk,
Russia; taimanov@math.nsc.ru}}
\title{Finite-gap solutions of the~modified Novikov--Veselov
equations: their spectral properties and applications
\thanks{Siberian Math. Journal {\bf 40} (1999), 1146--1156.}}
\date{}
\maketitle

\section{Introduction}

In~this article we construct finite-gap solutions of the~modified
Novikov--Veselov equations and discuss the~algebro-geometric
properties of the~corresponding spectral problem and its connection
with the~solutions of the~modified Korteweg--de Vries equations.
The~article adheres to [1] wherein finite-gap potentials
were written down (with a~short sketch of the~proof) for
a~two-dimensional Dirac operator.

The~modified Veselov--Novikov equations were introduced by
Bogdanov in [2] and have the~shape of a~Manakov ``L,A,B''-triple [3]:
$$
\frac{\partial L}{\partial t_n} = L A_n + B_n L,
                            \eqno{(1)}
$$
where
$$
L = \left(
\begin{array}{cc}
 0 & \partial \\ -\bar{\partial} & 0
\end{array}\right) + \left(\begin{array}{cc} U & 0 \\ 0 & U
\end{array}\right)
\eqno{(2)}
$$
and $A_n$ and $B_n$ are matrix differential operators; moreover,
the~order of $A_n$ equals $2n+1$.
Furthermore, the~highest term $A_n$ equals
$$
A_n =
\left(\begin{array}{cc} \partial^{2n+1} + \bar{\partial}^{2n+1}
& 0 \\ 0 & \partial^{2n+1} + \bar{\partial}^{2n+1}
\end{array}\right) + \dots\ .
\eqno{(3)}
$$

This hierarchy is adjoined to the~operator $L$
and preserves the~zero part of its spectrum. Indeed, if
$$
L \psi = 0
\eqno{(4)}
$$
then the~equation
$$
\frac{\partial
\psi}{\partial t_n} + A_n \psi = 0
\eqno{(5)}
$$
deforms the~eigenfunctions with the~zero eigenvalue; i.e.,
if (5) holds for all $t$ and if $\psi$ satisfies (4) for some $t_0$
then $\psi$ satisfies (4) for all $t$.

The~first equation ($n=1$) looks like
$$
U_t = \left(U_{zzz} + 3 U_z V + \frac{3}{2} U V_z \right) +
\left(U_{\bar{z}\bar{z}\bar{z}} + 3 U_{\bar{z}} \bar{V} +
\frac{3}{2} U \bar{V}_{\bar{z}}\right),
                                                        \eqno{(6)}
$$
where
$$
V_{\bar{z}} = (U^2)_z
$$
and $z = x + i y \in {\mathbb C}$. This equation
preserves realness of the~potentials $U$
and, for real potentials depending on a~single spatial variable $x$,
it is reduced to the~modified Korteweg--de Vries equation
$$
U_t =
\frac{1}{4} U_{xxx} + 6U_x U^2
$$
(here $V = U^2$).

The~second equation of the~modified Novikov--Veselov hierarchy
was written down in [4].

\section{Finite-gap potentials of a two-dimensional
Di\-rac operator and finite-gap solutions
of the modified Novikov--Veselov equations}

Like in [1], here we restrict exposition to the~case in which
the~spectral surface $\Gamma$ of the~operator $L$
is smooth. Even for the~operator corresponding to the~Clifford torus,
this surface is a~sphere with a~double point [5].
Explicit formulas for the~surfaces $\Gamma$ with double points
could also be constructed, but the~general construction
would be much more cumbersome. In~the most important case when
the~normalized surface remains connected, such solutions
appear in the~limit of the~solutions described below.

We~begin with recalling the~necessary results of [1]:
Propositions 1 and 2.

Consider the~more general operator
$$
\widetilde{L}=
\left(
\begin{array}{cc}
0 & \partial \\
-\bar{\partial} & 0
\end{array}
\right)
+
\left(
\begin{array}{cc}
U & 0 \\
0 & V
\end{array}
\right).
                                   \eqno{(7)}
$$
The~operator $\widetilde{L}$ is said to be {\it finite-gap} if
it has the~following shape.

\begin{proposition}
[\cite{1}]
Let~$\Gamma$ be a~compact Riemann surface of genus $g$,
let $\infty_{\pm}$ be
a~pair of distinct points in $\Gamma$;
let $k^{-1}_{\pm}$ be
local parameters in neighborhoods about these points; moreover,
$k^{-1}_{\pm}(\infty_{\pm}) =0$; and let $D$
a~nonspecial effective divisor of degree $g+1$ on $\Gamma \setminus
\{\infty_{\pm}\}$; i.e.,
$D=P_1 + \dots + P_{g+1}$, where
$P_i \in \Gamma \setminus \{\infty_{\pm}\}$.
Then

(1) There exists a~unique vector-function
$\psi(z,\bar{z},P) = (\psi_1, \psi_2)$,
where $z
\in {\mathbb C}$, such that $\psi$ is meromorphic in $P$
on $\Gamma \setminus \{\infty_{\pm}\}$,
has poles only at the~points of $D$,
and has the~following asymptotics:
$$
\psi = \exp{(k_+z)} \left[ \left(
\begin{array}{c} 1 \\ 0 \end{array} \right) + \left(
\begin{array}{c} \xi^+_{11}/k_+ \\ \xi^+_{21}/k_+ \end{array}
\right) + O\bigl(k_+^{-2}\bigr) \right] \ \
\mbox{as $P \to
\infty_+$},
$$
$$
\psi = \exp{(k_- \bar{z})} \left[ \left(
\begin{array}{c} 0 \\ 1  \end{array}\right) + \left(
\begin{array}{c} \xi^-_{11}/k_- \\ \xi^-_{21}/k_- \end{array}
\right) + O(k_-^{-2}) \right] \ \  \mbox{as $P \to
\infty_-$};
$$

(2) there exists a~unique operator $\widetilde{L}$
of the~shape (7) such that $\widetilde{L} \psi = 0$.
The~potentials of $\widetilde{L}$ have the~shape
$$
U =
-\xi^+_{21}, \ \ \ V = \xi^-_{11}.
\eqno{(8)}
$$
\end{proposition}

The~proof, implied in [1], bases on the~general properties
of the~Baker--Akhiezer functions [6,7].

Consider the~space
${\cal  E}_z = {\cal  E}(\Gamma,\infty_{\pm},k_{\pm},D,z)$
formed by the~functions $\varphi$ on $\Gamma$
satisfying the~following conditions:

(1) $\varphi$ is meromorphic outside $\infty_{\pm}$
and has poles only at the~points of $D$;

(2) $\varphi^+=\varphi \exp(-k_+ z)$ is holomorphic in
a~neighborhood of $\infty_+$
and  $\varphi^-=\varphi \exp(-k_- \bar{z})$
is holomorphic in a~neighborhood of $\infty_-$.

If~$\varphi_1, \varphi_2 \in {\cal  E}_z$ then the~function
$\varphi_1/\varphi_2$ is meromorphic on the~whole surface $\Gamma$
and, for $z$ in general position, the~divisor
of its poles is nonspecial. This implies for a~general $z$ that

(1) by the Riemann--Roch theorem,
$\dim {\cal  E}_z = 2$;

(2) $\varphi$ is uniquely determined by the~values
$\varphi^+(\infty_+)$ and $\varphi^-(\infty_-)$.

Now, take a~basis $\psi_1, \psi_2$ for ${\cal  E}_z$
normalized by the~conditions
$$
\psi_1 \exp(-k_+ z) = 1, \quad  \psi_2 \exp(-k_+ z)
= 0 \quad  \mbox{at $\infty_+$},
$$
$$
\psi_2 \exp(-k_- \bar{z}) = 0, \quad  \psi_2
\exp(-k_- \bar{z}) = 1 \quad \mbox{at $\infty_-$}.
$$
These basis functions have analytic continuations in $z$
on the~whole complex plane ${\mathbb C}$ to functions
defining the~components of a~Baker--Akhiezer function
$\psi = (\psi_1,\psi_2)$ satisfying the~conditions
of the~proposition.

Furthermore, the~components of $\psi$ give normalized
bases for ${\cal  E}_z$ which are unique.
This yields uniqueness of $\psi$ and proves
the~first claim of the~proposition.

If~the potentials of $L$ are defined by (8) then
$L\psi \in {\cal  E}_z$ for every $z$,
$\exp(-k_+ z)$ $L\psi = 0$ at $\infty_+$, and
$\exp(-k_-\bar{z})L\psi = 0$ at $\infty_-$.
The~last two equalities mean that $L\psi = 0$ everywhere.

The~proof of Proposition 1 is over.

\noindent
{\bf Remark 1.}
In~the case when $\Gamma$ has $m$ double points
exhausting all singularities, the~construction
of $\psi$ reduces to the~following: Let $\widetilde{\Gamma}$
be the~normalization of $\Gamma$ obtained by ``unsticking''
the double points, and under the~projection
$\widetilde{\Gamma} \to \Gamma$ the~pairs of points
$(Q^+_1,Q^-_1),\dots,(Q^+_m,Q^-_m)$
go into the~$m$ double points. Then $\psi$ is a~Baker--Akhiezer
function with the~same asymptotics but it has the~divisor  of poles $D$
of degree $g+m+1$ and satisfies the~normalization conditions
$\psi(Q^+_j) = \psi(Q^-_j)$, $j=1,\dots,m$.

\begin{proposition}
[\cite{1}]
Suppose that the~spectral data $(\Gamma,\infty_{\pm},k_{\pm},D)$
of a~fi\-ni\-te-gap operator $\widetilde{L}$ satisfy the~following
conditions:

(1) there are a~holomorphic involution
$\sigma:\Gamma \to
\Gamma$ such that
$\sigma(\infty_{\pm}) = \infty_{\pm}$,
$\sigma(k_{\pm}) = -k_{\pm}$,
and a~meromorphic differential
$\omega$ on $\Gamma$ with zeros in $D  + \sigma(D)$,
two poles at $\infty_{\pm}$, and principal parts
$\bigl(\pm k_{\pm}^2 + O\bigl(k_{\pm}^{-1}\bigr)\bigr) d  k_{\pm}^{-1}$;

(2) there are an~antiholomorphic involution
$\tau:\Gamma \to
\Gamma$ such that
$\tau(\infty_{\pm})$ $= \infty_{\mp}$,
$\tau(k_{\pm}) = -\bar{k}_{\mp}$,
and a~meromorphic differential
$\tilde{\omega}$ on $\Gamma$ with zeros in $D +
\tau(D)$, two poles at $\infty_{\pm}$,
and principal parts
$\bigl(k_{\pm}^2 + O\bigl(k_{\pm}^{-1}\bigr)\bigr) d k_{\pm}^{-1}$.

Then the~operator $\widetilde{L}$ has the~shape (2)
with a~real potential $U$.
\end{proposition}

Now, we are in a~position to formulate a~theorem
about finite-gap solutions of the~modified
Novikov--Veselov equations.

\begin{theorem}
Suppose that the~spectral data $(\Gamma,\infty_{\pm},k_{\pm},D)$
satisfy the~conditions of Propositions ~1 and 2. Then

(1) there is a~unique vector-function
$\psi(z,\bar{z},t_1,\dots,P)$ such that

(1a) $\psi$  depends on $z\in {\mathbb C}$, time variables
$t_1,\dots$, only finitely many of which  may differ from zero,
and $P \in \Gamma$;

(1b) $\psi$ is meromorphic in $P$ on $\Gamma \setminus \{\infty_{\pm}\}$
and has poles only at the~points of $D$;

(1c) the~following asymptotics hold:
$$
\psi = \exp\bigl(k_+ z + k_+^3 t_1 + \dots + k_+^{2n+1}t_n + \dots\bigr)
\left[\left(\begin{array}{c} 1 \\ 0 \end{array}\right) + O(k_+^{-1})
\right] \ \ \mbox{as $P \to \infty_+$},
$$
$$
\psi = \exp\bigl(k_- \bar{z} + k_-^3 t_1 + \dots + k_-^{2n+1}t_n + \dots\bigr)
\left[\left(\begin{array}{c} 0 \\ 1 \end{array}\right) + O(k_-^{-1})
\right] \ \ \mbox{as $P \to \infty_-$};
$$

(2) there are a~unique operator $L$ of the~shape (2) such that
$L\psi = 0$ and unique operators $A_n$ with principal parts
(3) such that equations (5) are satisfied;

(3) the~potential $U = U(z,\bar{z},t_1,\dots)$ of the~operator $L$
satisfies the~modified Novikov--Veselov equations (1).
\end{theorem}

Existence and uniqueness of $\psi$, $L$, and $A_n$
are proved in the~same way as in Proposition 1.
This implies that
$$
\frac{\partial (L\psi)}{\partial t_n} = \frac{\partial
L}{\partial t_n}\psi + L \frac{\partial \psi}{\partial t_n} =
\left( \frac{\partial L}{\partial t_n} - L A_n\right)\psi = 0.
$$
Proceeding as in the~derivation of the~Novikov--Veselov equations [8],
we now calculate some operators $B_n$ such that
$L A_n + B_n L$ are the~operators of multiplication by a~matrix,
i.e., matrix differential operators of zero order. For~example, for $n=1$
$$
B_1 = 3
\left(\begin{array}{cc} 0 & U_z \partial +
U_{\bar{z}}\bar{\partial} \\ -U_z \partial
-U_{\bar{z}}\bar{\partial} & 0 \end{array}\right)
+
$$
$$
+3\left(\begin{array}{cc} 0 & U_{\bar{z}\bar{z}} + U(\bar{V}-V) \\
-U_{zz}+U(\bar{V}-V) & 0
\end{array}\right).
$$
Define ${\cal  E}_{z,{\bf t}}$ as the~space of functions
$\varphi$ that are meromorphic on $\Gamma \setminus \{\infty_{\pm}\}$,
have poles only at the~points of $D$, and are such that
$\varphi^+ = \varphi \exp(-k_+ z - k_+^3 t_1 - \dots)$
is holomorphic in a~neighborhood of $\infty_+$
and $\varphi^- = \varphi \exp(-k_- \bar{z} - k_-^3 t_1 - \dots)$
is holomorphic in a~neighborhood of $\infty_-$.
For~$z,t_1,\dots$ in general position, these spaces
are two-dimensional and their elements are uniquely determined
by the~values of $\varphi^{\pm}$ at $\infty_{\pm}$.
The~multiplication
$$
\times(L_{t_n} - L A_n - B_n L):
{\cal  E}_{z,{\bf t}}
\to {\cal  E}_{z,{\bf t}}
                            \eqno{(9)}
$$
by a~matrix independent of $P \in \Gamma$
carries $\psi$ to the~vector-function $\psi_{(n)}$
whose components $\psi_{j,n}$ belong to ${\cal  E}_{z,{\bf t}}$
for arbitrary $z,t_1,\dots$.
By~the definition of $L$, $A_n$, and $B_n$, we have
$$
\psi_{j,n}^+(\infty_+) = \psi_{j,n}^-(\infty_-) = 0
$$
for $j=1,2$. It~follows that $\psi_{(n)} =0$ and (9)
is the~multiplication by the~zero $(2 \times 2)$-matrix.
Hence, equations (1) are satisfied, which completes the~proof of
Theorem ~1.

\begin{theorem}
Suppose that the~spectral data satisfy the~conditions
of Theorem ~1; moreover, there exists a~meromorphic function
$$
\lambda:\Gamma \to {\mathbb C} P^1 = \bar{{\mathbb C}},
$$
having exactly two poles at the~points $\infty_{\pm}$
with the~Laurent parts
$$
\lambda = \pm i k_{\pm} + O(1) \ \mbox{as}\ \ k_{\pm}
\to \infty.
$$
Then the~function $\psi(z,\bar{z},P)$ has the~shape
$$
\psi(z,\bar{z},P) = \tilde{\psi}(x,P) \exp{(\lambda(P)y)},
$$
the~potential of the~operator $L$ depends only on $x$
and its dynamics in times $t_n$ is described by the~equations
of the~modified Korteweg--de Vries hierarchy.
\end{theorem}

{\sl Proof}
of the~theorem reduces to the~following.
According to the~theory of Baker--Akhiezer functions,
there exists a~unique function $\tilde{\psi}(x,P)$ with the~prescribed
spectral data $(\Gamma,\infty_{\pm},k_{\pm},D)$
and the~asymptotics
$$
\tilde{\psi} \approx
\exp{(k_+ x)} \left( \begin{array}{c} 1 \\ 0 \end{array} \right)
\ \ \mbox{as $k_+ \to \infty$}, \ \
\tilde{\psi} \approx
\exp{(k_- x)} \left( \begin{array}{c} 0 \\ 1 \end{array} \right)
\ \ \mbox{as $k_- \to \infty$}.
$$
Afterwards, it is easy to see that the~function
$\tilde{\psi}(x,P) \exp{(\lambda(P)y)}$ satisfies
the~conditions of Theorem ~1 and so from uniqueness
we infer its coincidence with $\psi(z,\bar{z},P)$.

The~potential of the~operator $L$ is determined by the~function
$\tilde{\psi}$; in consequence, it depends only on $x$.
Moreover, its deformations in $t_n$ constitute
the~hierarchy of equations associated
with the~one-dimensional Dirac operator
$$
\widetilde{L} = \frac{1}{2}\left(\begin{array}{cc} 0 & \partial_x \\
-\partial_x & 0 \end{array}\right) +
\left(\begin{array}{cc} U & 0 \\ 0 & U \end{array}\right),
$$
and the~spectral problem (4) is reduced to the~Zakharov--Shabat problem
$$
\left[\left(\begin{array}{cc} 0 & \partial_x \\
-\partial_x & 0 \end{array}\right) +
\left(\begin{array}{cc} 2U & 0 \\ 0 & 2U
\end{array}\right)\right]\psi =
\left(\begin{array}{cc}
0 & i\lambda \\
i\lambda & 0 \end{array}\right)\psi.
                            \eqno{(10)}
$$
It~is easy to see that the~arising equations
constitute the~modified Korteweg--de Vries hierarchy.

\noindent
{\bf Remark 2.}
Under the~conditions of Theorem 2, the~Riemann surface $\Gamma$
is hyperelliptic, since on it there is a~meromorphic function
with exactly two poles.

Theorem 2, together with the~formulas of {\S}\,3,
provides derivation of explicit formulas for finite-gap
solutions of the~modified Korteweg--de Vries hierarchy.
In~another way (using the~Miura transformation),
this was made in [9].

\section{Explicit formulas for potentials and solutions}

Assume given spectral data
$(\Gamma,\infty_{\pm},k_{\pm},D)$
satisfying the~conditions of Theorem 1.

On~the Riemann surface $\Gamma$ of genus $g$, choose
a~canonical basis of 1-cycles: $\alpha_1,\dots,\alpha_g,
\beta_1,\dots,\beta_g$.
By~definition, its intersection form is as follows:
$$
\alpha_j \circ \beta_k = \delta_{jk}, \ \
\alpha_j \circ \alpha_k = \beta_j \circ \beta_k = 0.
$$
Given the~basis, we construct

(1)~the normalized basis of holomorphic 1-forms $\omega_1, \dots, \omega_g$:
$$
\int\limits_{\alpha_k} \omega_j = \delta_{jk};
$$

(2) the~matrix of $\beta$-periods of holomorphic 1-forms:
$$
\Omega_{jk} = \int\limits_{\beta_k} \omega_j;
$$

(3) the~theta function of the~surface $\Gamma$:
$$
\vartheta(u) = \sum_{N \in {\mathbb Z}^g} \exp{\pi i ((\Omega N,N) +2(N,u))},
$$
where $u \in {\mathbb C}^g$.

The~complex torus
$J(\Gamma) = {\mathbb C}^g / \{M + \Omega N : M,N \in {\mathbb Z}^g\}$
is called the~{\it Jacobian variety\/} of $\Gamma$ and the~mapping
$$
P \to  A(P) = \left(\int\limits_{P_0}^P \alpha_1,
\dots, \int\limits_{P_0}^P \alpha_g\right)
$$
from $\Gamma$ into $J(\Gamma)$ is called the~{\it Abelian mapping}.
Here $P_0$ is a~fixed point in $\Gamma$. By~linearity, the~Abelian mapping
extends to divisors, and the~expression $A(D_1) - A(D_2)$
is soundly defined for effective divisors of the~same degree and is
independent of the~choice of $P_0$ in the~definition of $A$.

Denote by $Q$ and $R$ effective divisors of degree $g$ such that
the~following relations hold:
$$
A(Q) + A(\infty_-) - A(D) = A(R) + A(\infty_+) - A(D) = 0.
$$
This amounts to fulfilment of the~linear equivalences
$$
D = P_1 + \dots + P_{g+1} \ \ \sim  \ \
Q_1 + \dots + Q_g + \infty_-,
$$
$$
D = P_1 + \dots + P_{g+1} \ \ \sim  \ \
R_1 + \dots + R_g + \infty_+,
$$
where $Q = Q_1 + \dots + Q_g$ and $R = R_1 + \dots + R_g$.

Denote by $\eta_l^{\pm}$ the~meromorphic 1-forms that are uniquely
determined by the~following conditions:

(1) $\eta_l^{\pm}$ has the~only pole at $\infty_{\pm}$
with the~Laurent part
$d k_{\pm}^l$;

(2) the~integrals of $\eta_l^{\pm}$   vanish over the~$\alpha$-cycles.

\noindent
With each form $\eta^{\pm}_l$, associate the~vector of $\beta$-periods
$$
(U^{\pm}_l)^j =
\frac{1}{2\pi i } \int\limits_{\beta_j} \eta_l^{\pm}
$$
and the~constants $a^{\pm}_l$ and $b^{\pm}_l$ defined by the~conditions
$$
\int\limits_{P_0}^P \eta^{\pm}_l - a^{\pm}_l = k^l_{\pm} + O\bigl(k^{-1}_{\pm}\bigr)
\ \mbox{near $\infty_{\pm}$}, \quad
\int\limits_{P_0}^P \eta^{\pm}_l - b^{\pm}_l = O\bigl(k^{-1}_{\mp}\bigr)
\ \mbox{near $\infty_{\mp}$}.
                                                                \eqno{(11)}
$$
Here we consider the~same paths from $P_0$ to $P$
for all $l$; i.e., defining asymptotics, we fix some
homotopic class of paths from $P_0$ to small neighborhoods
of the~infinities.

Denote by $\delta$ the~vector of Riemann constants
which is defined as follows: for a~point
$u \in J(\Gamma)$ in general position, the~function
$\vartheta(A(P)- u)$
has zeros  at exactly $g$ points $S_1, \dots, S_g$; moreover,
$u + \delta =
A(S_1) + \dots + A(S_g)$ (this function is multiple valued
and its values are determined by the~choice of the~integration path
in the~definition of the~Abelian mapping and differ by nonzero multipliers).

Also, choose an~odd half-period $\varepsilon \in J(\Gamma)$; i.e.,
$\vartheta(\varepsilon) = 0$ and $2\varepsilon \equiv 0$
on $J(\Gamma)$.

Existence and uniqueness, if any, of all the above-indicated objects
are well known from the~theory of Riemann surfaces (see, for
instance, [10]).

Define the~following functions:
$$
\Phi_1(z,{\bf t}) =
z
\left(\int\limits_{P_0}^P \eta^+_1 - a^+_1 \right) +
\bar{z}\left(\int\limits_{P_0}^P \eta^-_1 - b^-_1\right)
+
$$
$$
+\sum_{l \geq 1}
t_l \left(\int\limits_{P_0}^P \bigl(\eta^+_{2l+1} +
\eta^-_{2l+1}\bigr) - \bigl(a^+_{2l+1}+ b^-_{2l+1}\bigr)\right),
$$
$$
\Phi_2(z,{\bf t}) =
z\left(\int\limits_{P_0}^P
\eta^+_1 - b^+_1\right) +
\bar{z}\left(\int\limits_{P_0}^P \eta^-_1 - a^-_1\right)
+
$$
$$
+\sum_{l \geq 1}
t_l \left(\int\limits_{P_0}^P \bigl(\eta^+_{2l+1} + \eta^-_{2l+1}\bigr) -
\bigl(a^-_{2l+1}+ b^+_{2l+1}\bigr)\right),
$$
$$
\Psi(z,{\bf t}) = z\bigl(a^+_1 - b^+_1\bigr) +
\bar{z}\bigl(b^-_1 - a^-_1\bigr) +
\sum_{l \geq 1} t_l\bigl(a^+_{2l+1} - a^-_{2l+1} +
b^-_{2l+1}- b^+_{2l+1}\bigr),
$$
$$
F_1(z,{\bf t}) = U^+_1 z + U^-_1\bar{z}+
\sum_{l \geq 1}\bigl(U^+_{2l+1}+U^-_{2l+1}\bigr)t_l + \delta -
A(Q),
$$
$$
F_2(z,{\bf t}) = U^+_1 z + U^-_1\bar{z}+
\sum_{l \geq 1}\bigl(U^+_{2l+1}+U^-_{2l+1}\bigr)t_l + \delta -
A(R).
$$
To~simplify the~notation of the~arguments of theta functions,
denote the~values of the~Abelian mapping on some divisor (in particular,
at a~point) $S$ by $S$ rather than $A(S)$.

\begin{theorem}
The~function $\psi$ of Proposition 1 has the~following shape:
$$
\psi_1(z,{\bf t},P) =
\exp{(\Phi_1(z,{\bf t}))}
\cdot
\frac{\vartheta(P + F_1(z,{\bf t}))}
{\vartheta(P + \delta - Q)} \cdot
\frac{\vartheta(\infty_+ + \delta - Q)}
{\vartheta(\infty_+ + F_1(z,{\bf t}))} \times
$$
$$
\times
\frac{\vartheta(\varepsilon + P - \infty_-)}
{\vartheta(\varepsilon + \infty_+ -\infty_-)}
\cdot
\frac{
\prod^{g+1}\vartheta(\varepsilon + \infty_+ - P_j) \cdot
\prod^g\vartheta(\varepsilon + P - Q_j)}
{\prod^{g+1}\vartheta(\varepsilon + P - P_j) \cdot
\prod^g\vartheta(\varepsilon + \infty_+ - Q_j)},
$$
$$
\psi_2(z,{\bf t},P) =
\exp{(\Phi_2(z,{\bf t}))}
\cdot
\frac{\vartheta(P + F_2(z,{\bf t}))}
{\vartheta(P + \delta - R)} \cdot
\frac{\vartheta(\infty_- + \delta - R)}
{\vartheta(\infty_- + F_2(z,{\bf t}))} \times
$$
$$
\times
\frac{\vartheta(\varepsilon + P - \infty_+)}
{\vartheta(\varepsilon + \infty_- -\infty_+)}
\cdot
\frac{\prod^{g+1}\vartheta(\varepsilon + \infty_- - P_j) \cdot
\prod^g\vartheta(\varepsilon + P - R_j)}
{\prod^{g+1}\vartheta(\varepsilon + P - P_j) \cdot
\prod^g\vartheta(\varepsilon + \infty_- - R_j)}.
$$
The~potential $U$ has the~shape
$$
U(z,{\bf t}) =
- C \exp{(\Psi(z,{\bf t}))}
\frac{\vartheta(\infty_++ F_2(z,{\bf t}))}
{\vartheta(\infty_-+ F_2(z,{\bf t}))},
$$
where
$$
C =
\frac{\prod^{g+1}\vartheta(\varepsilon + \infty_- - P_j)
\prod^g \vartheta(\varepsilon + \infty_+ - R_j)}
{\prod^{g+1} \vartheta(\varepsilon + \infty_+ - P_j)
\prod^g\vartheta(\varepsilon + \infty_- - R_j)}
\times
$$
$$
\times
\frac{1}{\vartheta(\varepsilon + \infty_- \infty_+)}
\cdot
\sum (U^+_1)^j \frac{\partial \vartheta (\varepsilon)}{\partial
u^j}.
$$
\end{theorem}

Here we assume that in the~definition of the~Abelian mapping
at a~point $P$ we use the~same paths from $P_0$ to $P$
(because $P$ has several occurrences in the~same formula)
and the~paths joining $P_0$ with small neighborhoods
of the~points $\infty_{\pm}$ coincide with those
in the~definition of the~constants (11).

The~formulas for $\psi$ are verified directly by using
the~properties of theta functions [10,\,11].
For~the one-dimensional reduction (when the~potential $U$
depends only on a~single spatial variable) these formulas
were derived in [12] in detail. To~derive these formulas for $U$,
it suffices to use the~fact that $\partial
A(P)/\partial k^{-1}_+ = U^+_1$ at $\infty_+$.

\noindent
{\bf Remark 3.}
The~formulas for $\psi$ and $U$ can be simplified if we account
for the~fact that all differentials $\eta^{\pm}_{2m+1}$ for $m \geq 0$
are anti-invariant under the~involution $\sigma$.
This implies that $\psi$ and $U$ can be written down via
the~theta functions of the~Prym variety of the~covering
$\Gamma \to
\Gamma/\sigma$.
However, as compared with the~case of a~two-dimensional Schr{\"o}dinger
operator, which is finite-gap at a~single energy level [8,\,13],
here we cannot control the~topological type of the~involution
(in the~case of a~Schr{\"o}dinger operator the~analogous involution
has only two smooth fixed points and the~dimension of the~Prym
variety is half of the~dimension of $J(\Gamma)$).
We~merely recall a~simple inequality
for the~dimensions of the~Jacobian
and Prym varieties:
$$
\genus  (\Gamma) -
\genus (\Gamma/\sigma) = \dim \Prym (\Gamma,\sigma)
\geq \left[\frac{\genus (\Gamma)}{2}\right].
\eqno{(12)}
$$

\section{The~Floquet spectrum}

The~Riemann surface
$\Gamma$ appears naturally for the~operators with periodic coefficients;
namely, it is the~Floquet spectrum ``at the~zero energy level $E$.''

Assume that the~potentials $U$ and $V$ of the~operator (7)
are periodic with respect to a~lattice $\Lambda \subset {\mathbb C}$
of rank 2.
A~function $\psi: {\mathbb C} \to {\mathbb C}$
is said to be a~Floquet function of $L$ with eigenvalue $E$
and quasi-impulses $(k_1, k_2)$ if
$$
L \psi = E\psi ,\quad
\psi(z + \gamma) = \exp{(2\pi i (\Re \gamma \cdot k_1
+ \Im \gamma \cdot k_2) )}\psi(z)
$$
for $\gamma \in \Lambda$.
Each such function admits a~representation of the~shape
$$
\psi(z) = \exp{(2\pi i(xk_1 + yk_2))}\varphi(z),
$$
where $\varphi(z)$ is periodic with respect to $\Lambda$
and satisfies the~equation
$$
L_k \varphi = E \varphi,
\eqno{(13)}
$$
with
$$
L_k =
\left(
\begin{array}{cc}
0 & \partial \\
- \bar{\partial} & 0
\end{array}
\right)
+
\left(
\begin{array}{cc}
U &  \pi(k_2 + i k_1)
\\
\pi(k_2 - i k_1) & V
\end{array}
\right),
$$
and $\varphi$ can be regarded as a~function on
the~two-dimensional torus ${\mathbb C}/\Lambda$.
Choosing a~constant $C$ so that the~operator
$$
{\cal  A} = \left(
\begin{array}{cc}
C & \partial \\
- \bar{\partial} & C
\end{array}
\right)
$$
be invertible on $L_2({\mathbb C}/\Lambda)$, rewrite (13)
for $\xi =
{\cal  A}\varphi$ and obtain
$$
\left[
1 + \left( \begin{array}{cc} U -
(C+E)  & \pi(k_2 +  i k_1) \\ \pi(k_2 - i k_1) & V - (C+E)
\end{array}\right)
{\cal  A}^{-1}
\right] \xi = 0.
$$
The~last equation has the~shape
$$
(1 + A(k_1,k_2,E))\xi = 0,
                                                        \eqno{(14)}
$$
where $A(k_1,k_2,E)$ is a~polynomial pencil  in $k_1,k_2,E$
of compact operators from
$L_2({\mathbb C}/\Lambda)$
into $L_2({\mathbb C}/\Lambda)$.
Now, the~polynomial Fredholm alternative, first established by Keldysh [14],
implies that equation (14) is solvable if and only if
$(k_1,k_2,E)$ belongs to some complex-analytic submanifold of positive
codimension in ${\mathbb C}^3$.
The~same alternative implies that if $E=0$ then equation (14)
is solvable if and only if $(k_1,k_2)$ belongs
to the~complex-analytic submanifold $\widehat{\Gamma}$
of codimension 1 in ${\mathbb C}^2$ (see also [1]).

As~it is easy to see, the~manifold $\widehat{\Gamma}$
is invariant under the~action
of the~dual lattice
$\Lambda^{\ast}$:
$$
k_1 \to k_1 + \mbox{Re}\,\gamma^*,\quad  k_2 \to k_2 +
\mbox{Im}\,\gamma^{\ast},\quad  \gamma^{\ast}\in \Lambda^{\ast},
$$
where $\Lambda^{\ast}$ consists of the~vectors $\gamma^{\ast} \in {\mathbb C}$
such that
$(\gamma,\gamma^{\ast}) = \mbox{Re}\,\gamma \cdot
\mbox{Re}\,\gamma^{\ast} + \mbox{Im}\,\gamma \cdot
\mbox{Im}\,\gamma^{\ast} \in {\mathbb Z}$ for all
$\gamma \in \Lambda$.

The~factor-manifold $\Gamma = \widehat{\Gamma}/\Lambda^{\ast}$
is called the~{\it Floquet spectrum\/} of the~operator $L$
``at the~zero energy level.''

For~the twice periodic operators $L$ described in Proposition 1,
it is the~Riemann surface $\Gamma$ that represents the~Floquet spectrum.
In~this case the~function $\psi$ defines an~analytic family of
Floquet functions which has poles arising when we fix asymptotics.
A~rigorous proof of these facts can be carried out by
the~methods of perturbation theory and was implemented for
important two-dimensional scalar operators in [15].
We~should consider perturbations of the~zero potentials (in our case $U=V=0$)
for which the~structure of the~Floquet spectrum is rather simple.

\noindent
{\bf Example 1.}
$U = V = 0$.

We~may assume that $\Lambda = {\mathbb Z} + i {\mathbb Z}$.
The~Floquet functions are parametrized by two families
$\psi^1 = (e^{\lambda z},0)$ and $\psi^2 =(0,e^{\mu \bar{z}})$,
the~surface $\Gamma$ splits into a~$\lambda$-plane and a~$\mu$-plane,
the~$\lambda$-plane is compactified by the~point $\infty_+$,
and the~$\mu$-plane is compactified by the~point $\infty_-$.

\noindent
{\bf Example 2.}
$U = V = d \neq 0$.

We~again assume that $\Lambda = {\mathbb Z} + i {\mathbb Z}$.

The~Floquet functions are linear combinations of the~functions
$$
\tilde{\psi}(z,\bar{z},\lambda) = \left(
\exp{\left(\lambda z - \frac{d^2}{\lambda} \bar{z}\right)},
- \frac{d}{\lambda}\exp{\left(\lambda z - \frac{d^2}{\lambda}
\bar{z}\right)}\right),
$$
where
$\lambda \in {\mathbb C}^{\ast} = {\mathbb C} \setminus \{0\}$.
The~surface $\Gamma$ is the~complex sphere,
the~$\lambda$-plane, compactified by the~point at infinity.
Two~``infinities'' are distinguished on $\Gamma$:
$\infty_+$, where $\lambda=\infty$, and $\infty_-$, where
$\lambda=0$.
The~parameters $k_{\pm}$ have the~shape
$$
k_+ = \lambda, \ \ \ k_- = -\frac{d^2}{\lambda},
$$
the~divisor $D$ consists of the~point $d \in {\mathbb C}^{\ast}$,
and the~Baker--Akhiezer function $\psi$ has the~shape
$$
\psi =
\frac{\lambda}{\lambda - d}
\left(
\exp{\left(\lambda z - \frac{d^2}{\lambda} \bar{z}\right)},
- \frac{d}{\lambda}\exp{\left(\lambda z - \frac{d^2}{\lambda}
\bar{z}\right)}\right).
$$

\noindent
{\bf Example 3.}
$U = V$ is a~function in one variable.

Let~$U$ = $U(x)$ and $U(x+T) = U(x)$, where $T$ is a~minimal period.
Equation (4) is rewritten as the~Zakharov--Shabat system (10)
which in terms of $\eta_1 = \psi_1 + i\psi_2$
and $\eta_2 = \psi_1 - i\psi_2$
takes the~shape
$$
(\partial_x + 2iU) \eta_1 = -i \lambda \eta_2,\ \ \
(\partial_x - 2iU) \eta_2 = -i \lambda \eta_1,
$$
implying
$$
\bigl(\partial^2_x + 4U^2 + 2iU_x\bigr) \eta_1 = -\lambda^2 \eta_1,\ \ \
\bigl(\partial^2_x + 4U^2 - 2iU_x\bigr) \eta_2 = -\lambda^2 \eta_2.
$$
The~passage from the~operator
$\bigl(\partial^2_x + 4U^2 + 2iU_x\bigr) =
(\partial_x - 2iU)(\partial_x + 2iU)$
to the~operator
$\bigl(\partial^2_x + 4U^2 - 2iU_x\bigr) =
(\partial_x + 2iU)(\partial_x - 2iU)$
is referred to as the~{\it Miura transformation}.

According to the~theory of a~periodic one-dimensional Schr{\"o}dinger
operator [6], for an~operator of the~shape
$-\partial^2 + V(x)$ there is a~Floquet--Bloch function
$\tilde{\psi}(x,P)$
satisfying the~equation
$$
(-\partial^2 + V(x))\tilde{\psi} = E\tilde{\psi}
$$
and defined on a~two-fold covering of the~complex $E$-plane
of the~shape
$$
\widetilde{\Gamma} = \{ (\mu,E): \mu^2 = P(E)\}.
$$
Moreover, $P(E)$ is an~entire function which has only simple zeros and
is a~polynomial of odd degree when
the~number of these zeros is finite (i.e., when the~operator
is finite-gap), and $\mu$ is a~quasi-impulse in ~$x$.

Let~$L$ be a~finite-gap Dirac operator with
a~periodic one-dimensional potential $U(x)$.
Then the~Floquet spectrum of $L$ projects onto the~Floquet--Bloch
spectra of the~operators
$\bigl(\partial^2_x + 4U^2 + 2iU_x\bigr)$
and
$\bigl(\partial^2_x + 4U^2 -
2iU_x\bigr)$; moreover, this projection has the~simple shape
$(k_1,k_2)
\to \bigl(k_1,k_2^2\bigr)$.
Therefore, the~Floquet--Bloch spectra of two one-dimensional operators
connected by the~Miura transformation coincide (this was first proved
in [16]) and are the~factor-spaces $\Gamma/\omega$ with respect
to the~involution $\omega: (k_1,k_2) \to
(k_1,-k_2)$.  The~functions $\eta_1$
and $\eta_2$ are, up to multiplication by
meromorphic functions on $\Gamma$,
the~lifts to the~covering of the~Floquet--Bloch functions
of the~corresponding one-dimensional operators.

In~this language, Proposition 2 is interpreted as follows
(similar explanations of the~presence of involutions
of spectral surfaces were given in [17]
for many other matrix operators).

\begin{proposition}
Let~$L$ be a~twice periodic operator of the~shape (7).

1.~If $U = \bar{V}$ then
the~Floquet spectrum is preserved by the~anti-involution
$$
(k_1,k_2) \to (-\bar{k}_1,-\bar{k}_2);
$$

2.~If $U = \bar{U}$ and  $V = \bar{V}$ then
the~Floquet spectrum is preserved by the~anti-involution
$$
(k_1,k_2) \to (\bar{k}_1,\bar{k}_2).
$$
\end{proposition}

{\sl Proof}.
1.~Assume $U = \bar{V}$ and $L\psi = 0$, where $\psi =
(\psi_1,\psi_2)$ is a~Floquet function with quasi-impulses
$(k_1,k_2)$.  By~direct substitution we check that the~function
$\hat{\psi} =
(\bar{\psi}_2, - \bar{\psi}_1)$
is a~Floquet function with quasi-impulses
$(-\bar{k}_1,-\bar{k}_2)$.

2.~If $k = (k_1,k_2)$ belongs to the~Floquet spectrum then
$\dim \Ker L_k > 0$.
From the~explicit form of the~operators we see that
$L^{\ast}_k = L_{\bar{k}}$.
The~index of an~elliptic operator on the~compact manifold
${\mathbb C}/\Lambda$ is determined by its principal part;
in the~case of $L_k$ we have
$$
\ind L_k =
\dim \Ker L_k  - \dim \Ker L^{\ast}_k =
\ind\left(\begin{array}{cc} 0 & \partial \\ -\bar{\partial} & 0
\end{array}\right) = 0,
$$
since the~principal part is selfadjoint.
Therefore, the~inequality $\dim \Ker L_k >0$ implies
$\dim \Coker L_k \neq 0$
and $\dim L_{\bar{k}} > 0$, which completes the~proof
of Proposition 3.

In~terms of Proposition 2, the~anti-involution
$(k_1,k_2) \to (\bar{k}_1,\bar{k}_2)$
is $\tau$, and the~composite of two anti-involutions of Proposition 3
is the~involution $\sigma$.

\section{On~soliton deformations of tori}

According to [1], each two-dimensional torus $\Sigma$,
immersed in ${\mathbb R}^3$ and smooth of the~class $C^3$,
is represented by the~formulas
$$
X^1(z,\bar{z}) = X^1_0 +
\frac{i}{2} \int^z_0 \bigl(\bigl(\bar{\psi}_2^2 + \psi_1^2\bigr)\,
dz' - \bigl(\bar{\psi}_1^2 + \psi_2^2\bigr)\, d\bar{z}' \bigr),
$$
$$
X^2(z,\bar{z}) = X^2_0 + \frac{1}{2}\int^z_0
\bigl(\bigl (\bar{\psi}_2^2 - \psi_1^2)\, dz' -
\bigl(\bar{\psi}_1^2 - \psi_2^2\bigr)\,d\bar{z}' \bigr),
                                                                \eqno{(15)}
$$
$$
X^3(z,\bar{z}) = X^3_0 + \int^z_0
(\psi_1 \bar{\psi}_2 dz' + \bar{\psi}_1 \psi_2 d\bar{z}'),
$$
where $X_0 \in {\mathbb R}^3$
is a~point lying on the~torus and $\psi$
is a~Floquet function of the~operator (2) defined on ${\mathbb C}$
and periodic with respect to the~lattice $\Lambda$.
Moreover, the~function $\psi$ is multiplied by $\pm 1$ under
shifts by periods and the~torus $\Sigma$
is conformally equivalent to the~flat torus ${\mathbb C}/\Lambda$.

For~surfaces locally defined by formulas (15),
B.~G.~Konopelchenko introduced deformations that are
described by the~modified Novikov--Veselov equations (see [18]):
one should deform $U$ in accordance with the~modified
Novikov--Veselov equations, at that deforming the~function $\psi$
in accordance with~ (5).
The~deformation of $\psi$ gives rise to a~local deformation
of the~surface.

It~is these deformations that stimulated us to
present formulas for $\psi$ and its soliton deformations (see {\S}\,3).

It~was shown in [4] that if a~torus is given in advance then
this deformation, corresponding to ~(6), gives rise to a~global
deformation of the~torus; moreover, the~Willmore functional
(the integral of squared mean curvature) is a~first integral
$$
{\cal  W}(\Sigma) = 4 \int\limits_{{\mathbb C}/\Lambda}
U^2(z,\bar{z}) \,dx dy.
$$

The~well-known Willmore conjecture asserts that the~minimum of
the Willmore functional on immersed tori equals $4\pi^2$.
The~Willmore functional is invariant under the~conformal transformations
of ${\mathbb R}^3$ that do not send the~points of the~torus to
the~point at infinity.

Basing on the~fact that a~minimum of such variational problem
must be nondegenerate (after factorization by the~action of the~conformal
group), it was conjectured in [4] that the~minima of this functional
are stationary under the~deformations generated by equation (6).
We~can extend this as follows:
{\sl for all equations of the~modified Novikov--Veselov hierarchy,
the~minima of ${\cal  W}$ for fixed conformal classes
are stationary under the~induced deformations.}

According to this conjecture, for the~minima of the~Willmore functional,
the~Prym variety of the~covering $\Gamma \to \Gamma/\sigma$
must be one-dimensional and the~deformations generated by the~modified
Novikov--Veselov equations reduce to translations of tori
along themselves. According to (12), this implies
$\genus (\Gamma) \leq 3$.  For~the surfaces of genus $3$,
the~dimension of the~Prym variety equals $3$ if $\sigma$
is a~hyperelliptic involution and $2$ if $\sigma$
has $4$ fixed points (the case in which the~dimension equals 1
corresponds to an~involution without fixed points).
Consequently, the~last conjecture implies that for the~minima
we must have $\genus (\Gamma) \leq 2$.

\end{document}